\documentclass[12pt]{article}
\input{bayesuvius.sty}
\newcommand{\tilP}[0]{\tilde{P}}
\newcommand{\link }[0]{\text{---}}
\newcommand{\linkab }[0]{{\rva\text{---}\rvb}}

\begin{document}
\title{Goodness of Causal Fit}
\date{ \today}
\author{Robert R. Tucci\\
        tucci@ar-tiste.com}
\maketitle
\vskip2cm
\section*{Abstract}
We propose a 
Goodness of Causal Fit (GCF) measure
which depends 
on Judea Pearl's ``do" interventions.
This is different
from Goodness of Fit (GF) measures,
which do not use interventions.
Given a set ${\cal G}$
of DAGs with the same nodes,
to find a good $G\in {\cal G}$,
we propose plotting
$GCF(G)$ versus $GF(G)$
for all $G\in {\cal G}$,
and finding a 
graph $G\in {\cal G}$  with 
a large amount 
of both types of goodness.

\newpage
\section{Introduction}

Frequently,
when students
first encounter
Bayesian Networks (bnets)
and Causal Inference (CI)
(Refs.\cite{pearl-2013book},
\cite{bayesuvius}),
they experience serious doubts
about the usefulness of this
theory, because they believe
finding the underlying model 
(i.e., DAG)
for most realistic
 physical situations is
too difficult or impossible.
I believe
that part of the problem
is that these students
are assuming, perhaps
unconsciously,
that there exists
a unique DAG
that fits Nature perfectly,
and a mind-boggling number
 of possibilities
to sift through to find that DAG.
Rather than looking
for a unique DAG,
I think a better strategy
is to write down
a set $\calg$ 
of likely DAGs,
and to calculate for 
each DAG in $\calg$,
 a measure called 
Goodness of Causal Fit (GCF).
Then use a DAG with
 a high GCF score.
 
 The goal of this paper
 is to propose a GCF measure.
 Such a measure is of course
 not unique,
 and someone may propose 
 in the future a measure that is better
 than ours.

 It's clear that any measure
 of GCF will have to
 involve interventions
 such as the ``do" intervention
 (see 
 Refs. \cite{pearl-2013book}
 and \cite{bayesuvius})
 invented by Judea Pearl et al.
 Without interventions like ``do",
 it might be impossible
 to distinguish which DAG
 of a set is the best causal fit.
 For example, the family of triangular
 bnets can 
 all represent the same
 probability distribution
 because they are fully connected.
Hence, from the 
probability distribution of the
triangular bnet alone,
it is impossible to decide
which bnet in the family is
the best causal fit for the
physical situation 
being considered.

When designing a GCF measure,
it is important to keep
in mind the Data Axiom\footnote{
This is just my whimsical name for it.} of CI: A dataset is causal model-free.
 In the Data Axiom,
when we say a ``dataset", we are referring to  a table of data, where all
 the entries of each column have the same units, and 
measure a single feature, and each row refers to one
 particular sample or individual. Datasets are particularly 
useful for estimating probability distributions and for 
training neural nets. In the Data Axiom, when we say ``causal model", we are 
referring to a DAG (directed acyclic graph) or a bnet
 (bnet= DAG + probability table 
for each node of DAG).

You can try to derive a causal model from a dataset, 
but you'll soon find out that you can only go so far.
The process of finding a {\it partial} causal model from a dataset 
is called structure learning (SL).  SL can be done quite
 nicely with Marco Scutari's open source program 
{\tt bnlearn} (Ref\cite{bnlearn}).
The problem is that SL often cannot 
narrow down the causal model to a single one. It finds an undirected graph (UG), 
and it can determine the direction of some of the arrows in the UG, 
but it is often incapable, for well understood 
fundamental ---not just technical--- reasons,
 of finding the direction of {\it all}  the arrows of the UG. 
So it often fails to fully specify a DAG.

Let's call the ordered pair (dataset, causal model) a 
{\bf dataset++}.
 Then what the Data Axiom is saying is that a dataset 
is causal model-free or model-less (although sometimes one can 
find a partial causal model hidden in there). 
A dataset is not a dataset++.

Graphs
which contain both directed 
and undirected edges
are called 
{\bf partially directed (PD) graphs}.
{\tt bnlearn} takes
a dataset as input
and returns a PD graph
$G_{pd}$.
Given a PD graph $G_{pd}$,
let $\calg_{max}(G_{pd})$
be the DAG set 
which 
is generated
by giving directions to all 
undirected edges of $G_{pd}$
in all possible ways.
We will refer to the
DAG set
 $\calg_{max}(G_{pd})$ as the
 {\bf maximal generation of $G_{pd}$}
and to any subset $\calg(G_{pd})$ of
$\calg_{max}(G_{pd})$ as a {\bf non-maximal 
generation of $G_{pd}$.}
Once we define below our
 GCF measure,
we will evaluate it for the DAGs of 
non-maximal generation $\calg(G_{pg})$.

Henceforth, random variables
will be indicated by underlining.
Also, Pearl's do operator assignment $do(\rvx)=x$
will be denoted by $\cald\rvx =x$.
Both
of these notational
conventions are also used in Ref.\cite{bayesuvius}.

\section{Goodness of Fit}
Before trying to
define a GCF measure,
it
is instructive to review 
the closely related, well established, measures
of Goodness of Fit (GF).

Consider 
two
probability distributions
$PO(x)$ and $PE(x)$,
where $x\in S_\rvx$.
By a GF measure, we mean 
a measure of the 
difference between 
$PO$ and $PE$.
Usually $PO$ is the
observed probability distribution and 
$PE$ is the expected, theoretical one.

Three popular
measures of
the difference between $PO$ and $PE$
are:
\begin{enumerate}
\item
The
{\bf Kullback-Liebler divergence}:
\beq
D_{KL}(PO\parallel PE) =
\sum_{x\in S_\rvx}
PO(x)\ln \frac{PO(x)}{PE(x)}
\;.
\eeq
\item
The
{\bf Pearson divergence}
(a.k.a. {\bf Pearson Chi-squared test statistic}):
\beq
D_{\chi^2}(PO\parallel PE)=
\sum_{x\in S_\rvx}
\frac{[PO(x)-PE(x)]^2}{PE(x)}
=
\sum_{x\in S_\rvx}
\frac{PO^2(x)}{PE(x)}-1
\;.
\eeq

It's easy to show 
using $\ln(1+\delta)=\delta +\calo(\delta^2)$
that
if $\left|\frac{PO(x)}{PE(x)}-1\right|<<1$
for all $x\in S_\rvx$, then

\beq
D_{KL}(PO\parallel PE)\approx 
D_{\chi^2}(PO\parallel PE)
\eeq

\item
The {\bf Euclidean distance squared}:

\beq
D_E(PO,PE)=
\sum_{x\in S_\rvx}
[PO(x)-PE(x)]^2
\eeq
\end{enumerate}
Note that of these 3 measures,
only $D_E(PO, PE)$ is symmetric 
in $PO$ and $PE$.

Given any bnet $G$
with full probability
distribution
\footnote{We define
$x.$
to be a vector
with components $x_i$}
  $P_G(x.)$
and a
probability distribution\footnote{
Empirical distributions will 
be denoted by $P$ with a tilde over it.}
$\tilP(x.)$
derived empirically from a dataset,
let

\beqa
D(G)
&=&
\sum_{x.}
\tilP(x.)\ln 
\frac{\tilP(x.)}
{P_{G}(x.)}
\\
&=&
D_{KL}(
\tilP(\rvx.)\parallel P_{G}(\rvx.))
\eeqa
We define {\bf Goodness of Fit (GF)}
of the bnet $G$ by

\beq
GF(G)=\ln \frac{1}
{D(G)}
\eeq

\section{GCF example 1}

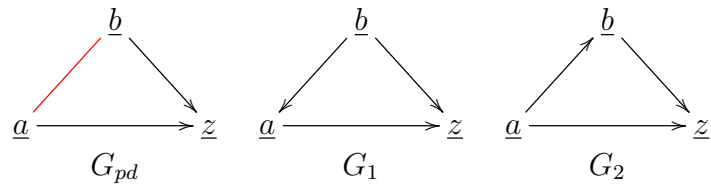
\begin{figure}[h!]
$$
\begin{array}{ccc}
\xymatrix{
&\rvb\ar@[red]@{-}[ld]\ar[dr]
\\
\rva\ar[rr]&&\rvz
}
&
\xymatrix{
&\rvb\ar[dl]\ar[dr]
\\
\rva\ar[rr]&&\rvz
}
&
\xymatrix{
&\rvb\ar[dr]
\\
\rva\ar[ur]\ar[rr]&&\rvz
}
\\
G_{pd}&G_1&G_2
\end{array}
$$
\caption{$\calg(G_{pd})=\{G_1, G_2\}$.
From the partially directed graph $G_{pd}$,
one can
generate the DAGs $G_1$ and $G_2$
by giving directions to
all undirected edges of $G_{pd}$
in
all possible ways.
(In this case, there is only one
undirected edge in $G_{pd}$.) }
\label{fig-ob-eq-1}
\end{figure}

For the first example of
our GCF measure,
we consider 
$\calg(G_{pd})=\{G_1, G_2\}$
given by Fig.\ref{fig-ob-eq-1}.
We will assume the following:

\begin{itemize}
\item
First, we assume that we have collected
a dataset from which we have
extracted a full empirical
distribution
$\tilP(z, a,b)$.
From $\tilP(z, a,b)$,
we assume that the following
have been calculated.
$\tilP(a)$, $\tilP(b)$.
\item
Second, we assume that a
dataset has been collected
 for which $\rva$ was held
fixed to each of
the possible values
$a\in S_\rva$ of $\rva$.
Furthermore, we assume
that the distribution
$\tilP(b|\cald \rva=a)$
has been calculated from that dataset.
\item
Third, we assume that a
dataset has been collected
 for which $\rvb$ was held
fixed to each of
the possible values
$b\in S_\rvb$ of $\rvb$.
Furthermore, we assume that
the distribution
$\tilP( a|\cald \rvb =b)$
has been calculated
from that dataset.
\end{itemize}
We will refer to
$\tilP(b|\cald \rva=a)$
and 
$\tilP(a|\cald \rvb =b)$
as {\bf empirical do-probability distributions}.

Now define
\beqa
\calh_a^\linkab&=&
\sum_{b}\tilP(b)
\ln
\frac
{\tilP(b)}
{\tilP(b|\cald \rva=a)}
\\
&=&D_{KL}(\tilP(\rvb)
\parallel 
\tilP(\rvb|\cald \rva=a))
\\
\calh_\rva^\linkab &=& \sum_a \tilP(a) \calh_a^\linkab
\\&=& E_a[\calh_a^\linkab]
\eeqa
and

\beqa
\calh_b^\linkab
&=&
D_{KL}(\tilP(\rva)
\parallel \tilP(\rva|\cald\rvb=b) )
\\
\calh_\rvb^\linkab 
&=&
\sum_a \tilP(b) \calh_b^\linkab
\\&=& E_b[\calh_b^\linkab]
\;.
\eeqa
We will
refer to $\calh_\rvx$ for any node $\rvx$
as the {\bf hospitality of node $\rvx$}.
Note that the 
hospitality for node 
$\rvx$ is zero
if node $\rvx$ has no incoming
arrows (i.e., 
is ``inhospitable"), and becomes 
positive if node $\rvx$
does have some incoming arrows
(i.e., is ``hospitable").

Note that 
if the truth is $G_2$ with $\rva\rarrow \rvb$,
then
\beq
\calh_a^\linkab=0
\text{ for all $a$ so }
\underbrace{\calh_\rva^\linkab}_0
\leq \calh_\rvb^\linkab
\eeq
and
if the truth is
 $G_1$ with $\rvb\rarrow \rva$, then

\beq
\calh_b^\linkab=0
\text{ for all $b$ so }
\calh_\rva^\linkab\geq 
\underbrace{\calh_\rvb^\linkab}_0
\;.
\eeq
Hence, 
no matter what the truth is, the arrow 
connecting nodes $\rva$
and $\rvb$ always points towards
the larger of the 2 hospitalities
(i.e., the arrow ``seeks the most hospitable
node")

If $\calh_\rva^\linkab\leq  \calh_\rvb^\linkab$, then define
$GCF(G_1)=-1$ and $GCF(G_2)=+1$.

If $\calh_\rvb^\linkab\leq  \calh_\rva^\linkab$, then define
$GCF(G_1)=+1$ and $GCF(G_2)=-1$.

\section{GCF example 2}

\begin{figure}[h!]
$$
\xymatrix{
&\rvx_1
\ar@[red]@{-}[dl]\ar@[red]@{-}[dr]
\\
\rvx_2\ar[dr]
&&\rvx_3\ar[dl]
\\
&\rvx_4\ar[d]
\\
&\rvx_5
\\
&G_{pd}
}
\;\;\;\;\;
\xymatrix{
&\rvx_1
\ar[dl]
\ar[dr]
\\
\rvx_2\ar[dr]
&&\rvx_3\ar[dl]
\\
&\rvx_4\ar[d]
\\
&\rvx_5
\\
&G_1
}
\;\;\;\;\;
\xymatrix{
&\rvx_1
\ar[dr]
\\
\rvx_2\ar[dr]\ar[ur]
&&\rvx_3\ar[dl]
\\
&\rvx_4\ar[d]
\\
&\rvx_5
\\
&G_2
}
\;\;\;\;\;
\xymatrix{
&\rvx_1
\ar[dl]
\\
\rvx_2\ar[dr]
&&\rvx_3\ar[dl]\ar[ul]
\\
&\rvx_4\ar[d]
\\
&\rvx_5
\\
&G_3
}
$$
\caption{$\calg=\{G_1, G_2, G_3\}$.
$\calg$ is a set of observationally
equivalent (OE) graphs. 
These are graphs that have the
same full
probability distribution, and
are therefore indistinguishable
by means of GF alone. For more info about 
OE graphs, see Chapter
entitled ``Observationally Equivalent DAGs"
in Ref.\cite{bayesuvius}.
Note that $\calg_{max}(G_{pd})$
includes one more DAG,
the one in which node $\rvx_1$
is a collider.
Hence $\calg$
is a non-maximal generation of $G_{pd}$.}
\label{fig-ob-eq-2}
\end{figure}
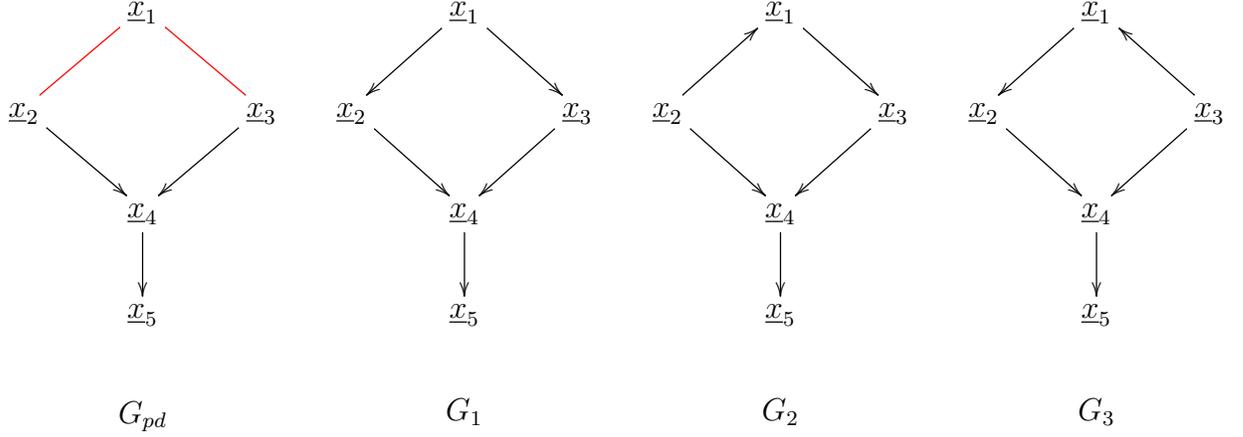

For the second example of
our GCF measure,
consider 
$\calg=\{G_1, G_2, G_3\}$
given by Fig.\ref{fig-ob-eq-2}.

The relative size of the hospitalities
$\calh_{\rvx_2}^{\rvx_2\link \rvx_1}$, 
$\calh_{\rvx_1}^{\rvx_2\link \rvx_1}$,
 $\calh_{\rvx_1}^{\rvx_1\link\rvx_3}$ and $\calh_{\rvx_3}^{\rvx_1\link\rvx_3}$,
depends on the empirical
do-probability distributions.
For definiteness,
suppose the
sizes of these hospitalities
are related as follows:

\beq
\calh_{\rvx_2}^{\rvx_2\link \rvx_1}\leq  \calh_{\rvx_1}^{\rvx_2\link \rvx_1}
,\quad\quad
 \calh_{\rvx_1}^{\rvx_1\link\rvx_3}
 \leq  \calh_{\rvx_3}^{\rvx_1\link\rvx_3}
\;.
\eeq

For any two hospitalities
$\calh_\rva^\linkab$ and $\calh_\rvb^\linkab$,
let
\beq
d_{\rvb,\rva}=
|\calh_\rvb^\linkab-\calh_\rva^\linkab|
\eeq

If we abbreviate $\rvx_j$ by $j$, 
we can define the GCF for each of
the graphs in $\calg$ by:

\begin{subequations}
\label{eq-gcf-example2}
\beq
GCF(G_1)=
\frac{
-d_{2,1} + d_{1,3}
}
{d_{2,1} + d_{1,3}}
\eeq

\beq
GCF(G_2)=
\frac{
d_{2,1} + d_{1,3}
}
{d_{2,1} + d_{1,3}}=1
\eeq

\beq
GCF(G_3)=
\frac{
-d_{2,1} - d_{1,3}
}
{d_{2,1} + d_{1,3}}=-1
\eeq
\end{subequations}

\section{GCF in general}
Suppose $G_i\in\calg$,
where $\calg$ is a non-maximal generation
of $G_{pd}$.
In that case,
we define a GCF measure
as follows.
Note that the
following definition generalizes
the definition of GCF measure 
that was used in the 2
special cases that
we have considered so far.

For any 
bnet 
$G_i\in\calg$
with nodes $\rva$ and $\rvb$,
define the {\bf hospitality of node $\rvb$} by

\beqa
\calh_b^\linkab
&=&
D_{KL}(\tilP(\rva)
\parallel \tilP(\rva|\cald\rvb=b))
\\
\calh_\rvb^\linkab 
&=&
\sum_a \tilP(b) \calh_b^\linkab
\\&=& E_b[\calh_b^\linkab]
\;.
\eeqa

For any two hospitalities
$\calh_\rva^\linkab$ and $\calh_\rvb^\linkab$,
define the {\bf hospitality distance} by
\beq
d_{\rvb,\rva}=
|\calh_\rvb^\linkab-\calh_\rva^\linkab|
\eeq
Note that $d_{\rvb,\rva}=0$
iff $\calh_\rva^\linkab=\calh_\rvb^\linkab$.
See Appendix A for a proof that
if $\calh_\rva^\linkab=\calh_\rvb^\linkab$, then
there is no arrow between
$\rva$ and $\rvb$.

For any $G_i\in\calg$,
define 
the {\bf edge reward function} 
 by
\beq
\rho_{G_i}(\rva \link \rvb)=
\left\{
\begin{array}{ll}
+1&\text{ if edge
$\rva\link \rvb$ in $G_i$
points
towards the larger of $\calh_\rva^\linkab$ and $\calh_\rvb^\linkab$.}
\\
-1&\text{ otherwise}
\end{array}
\right.
\eeq

Now suppose that
$\calg$
is either a maximal
or non-maximal
generation of
PD graph $G_{pd}$
with undirected edges
$\{\rva_k\link 
\rvb_k\}_{k=0,1, \ldots, nk-1}$.
Then 
define the GCF of 
graph
$G_i\in \calg$ by

\beq
GCF(G_i)= 
\frac{
\sum_{k=0}^{nk-1}
\rho_{G_i}(\rva_k\link \rvb_k)
d_{\rva_k, \rvb_k}
}
{
\sum_{k=0}^{nk-1}
d_{\rva_k, \rvb_k}
}
\;.
\label{eq-rel-gfc}
\eeq
Note that
$-1\leq GCF(G_i)  \leq 1$.

If the DAG set $\calg$ 
contains only one DAG $G$,
define $GCF(G)=1$, because the
directions of 
all arrows in $G$ are known.

Call an undirected graph a {\bf frame}
and define the
{\bf frame of a DAG}
to be the frame that one obtains
by turning
all the edges of the DAG from directed 
to undirected ones.

So far, we 
have applied our GCF measure
 to a DAG set $\calg$
which is
either a maximal or 
non-maximal generation
of  a PD graph $G_{pd}$,
or is a singleton set.
But what if we want a GCF
that can score every DAG
in a  DAG set
$\calg$ that contains
DAGs with different frames
but the same nodes?
In that case, 
let $F$
be the frame
which 
is the union of 
all edges in all $G\in \calg$.
For each edge $\rva\link \rvb$ of $F$,
if all the $G\in \calg$
give the same direction
to that edge, then give that direction
to that edge in $F$.
After doing this for
all edges of $F$,
call $G_{pd} $ the resulting 
PD graph.
Modify each $G\in \calg$
by adding to it the undirected edges
that occur in $G_{pd}$
but not in $G$.
The new $G$, call it $[G]_{mod}$,
is PD. Remove $G$ from $\calg$
and add to $\calg$
the elements of
the maximal generation $\calg_{max}([G]_{mod})$.
At this point,
we have reduced our
seemingly more 
complicated situation
where $\calg$ contains
different frames with the same nodes
to the original situation
in which $\calg$
is a non-maximal 
generation of $G_{pd}$.

So let $\calg$
be an arbitrary set of
DAGs with the same nodes.
Our GCF  measure
is not enough to
decide the best 
possible $G$ in $\calg$,
because there might 
be several graphs with 
$GCF\approx 1$.
For this reason,
we recommend
plotting $GCF(G)$ 
versus $GF(G)$
for all $G\in\calg$.
Then  choose a $G$ with a
large amount
of both types of goodness.
A plot of 
$GCF(G)$
versus $GF(G)$
agrees with the spirit of
the Data 
Axiom,
because in that axiom 
we also acknowledge a separation between the
degrees of freedom of the
dataset  and 
those of the causal model.

\appendix

\section{Appendix}

\begin{figure}[h!]
$$
\begin{array}{ccc}
\xymatrix{
&\rvn\ar[ld]\ar[rd]\ar[dd]
\\
\rva\ar[rd]
&&\rvb\ar[ll]
\ar[ld]
\\
&
\rvs
}
&\quad&
\xymatrix{
&\rvn\ar[rd]\ar[dd]
\\
\cald\rva = a\ar[rd]
&&\rvb
\ar[ld]
\\
&
\rvs
}
\\
\\
(A)&&\quad(B)
\end{array}
$$
\caption{Bnets used 
to prove Claim \ref{cl-ha-hb}.
The proof is also valid if the
direction of
arrow $\rvn\rarrow\rvs$
is reversed. }
\label{fig-ha-hb}
\end{figure}
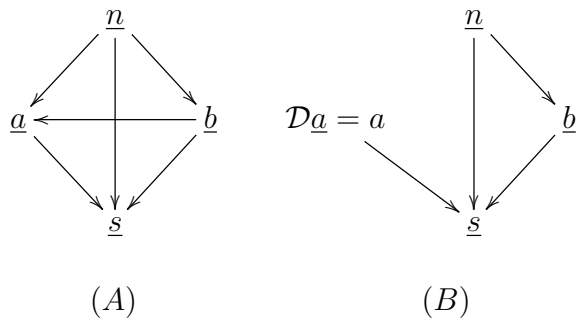

\begin{claim}\label{cl-ha-hb}
Suppose $\rva, \rvb$ 
are any two nodes of
a bnet $G_i$. Then
either 
$\calh_\rvb^\linkab=0$ or  $\calh_\rva^\linkab=0$.
\begin{enumerate}
\item 
If $\calh_\rvb^\linkab=0$ and  $\calh_\rva^\linkab\neq 0$, then
the arrow between $\rva$
and $\rvb$ points towards $\rva$
(i.e., towards large hospitality).
\item
If $\calh_\rvb^\linkab\neq 0$ and  $\calh_\rva^\linkab= 0$, then
the arrow between $\rva$
and $\rvb$ points towards $\rvb$
(i.e., towards large hospitality).

\item
If
$\calh_\rvb^\linkab= \calh_\rva^\linkab=0$,
then 
there is no arrow between
$\rva$ and $\rvb$.

\end{enumerate}
\end{claim}
\proof

Consider Fig.\ref{fig-ha-hb}.
In that figure, $\rvn$ and $\rvs$
might each represent multiple nodes
of $G_i$.
Note that in 
Fig.\ref{fig-ha-hb}(B),
all paths connecting nodes $\cald \rva=a$ and 
$\rvb$ are blocked by a collider
so these two nodes are independent
random variables.
Hence, $P(b|\cald \rva=a)= P(b)$
and $\calh_\rva^\linkab=0$.
If the labels $\rva$ and $\rvb$ are 
interchanged, then 
$\calh_\rvb^\linkab=0$.
If both hospitalities 
are zero, then there can't be any arrow
between $\rva$ and $\rvb$.

\begin{figure}[h!]
\centering
\includegraphics[width=3in]
{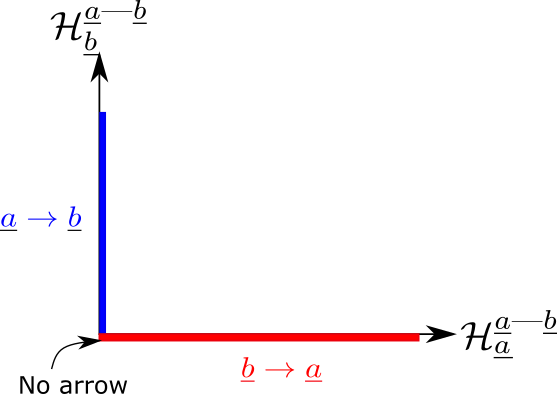}
\caption{Plot of 2 hospitalities 
for link $\linkab$.
All allowed values 
fall in the
red or blue regions.
If a point falls in
the blue region, then the arrow 
points from $\rva$ to $\rvb$,
and if it falls in the red region,
then
the arrow points from $\rvb$ to $\rva$.
If it falls at the origin, then
there is no arrow between 
nodes $\rva$ and $\rvb$. }
\label{fig-hospitalities-x-y}
\end{figure}

The results of this claim
are represented graphically in 
Fig.\ref{fig-hospitalities-x-y}
\qed

\bibliographystyle{plain}
\bibliography{references}

\begin{thebibliography}{1}

\bibitem{pearl-2013book}
Judea Pearl.
\newblock {\em Causality: Models, Reasoning, and Inference, Second Edition}.
\newblock Cambridge University Press, 2013.

\bibitem{bnlearn}
Marco Scutari.
\newblock bnlearn.
\newblock \url{https://www.bnlearn.com/}.

\bibitem{bayesuvius}
Robert~R. Tucci.
\newblock Bayesuvius (book).
\newblock \url{ https://github.com/rrtucci/Bayesuvius/raw/master/main.pdf}.

\end{thebibliography}
\end{document}